\documentclass[12pt]{article}
\usepackage{a4p}
\usepackage{epsfig}
\usepackage{amsmath}     

\begin{document}
%
%
\newcommand{\PPEnum}    {CERN-EP/99-036}
\newcommand{\Date}      {01 March 1999}
\newcommand{\Author}    {Shoji Asai and Sachio Komamiya}
\newcommand{\MailAddr}  {Shoji.Asai@@cern.ch}
\newcommand{\EdBoard}   {Benno List, Homer Neal, Hartmut Rick and \\
Claire Shepherd-Themistocleous}
\newcommand{\DraftVer}  {Version 5.1}
\newcommand{\DraftDate} {25 Feb. 1999}
\newcommand{\TimeLimit} {25 Feb. 1999; 00:00 h (CERN TIME)}
\newcommand{\Submit} {To be submitted to Phys. Lett. B}

%
 
\def\toprule{\noalign{\hrule \medskip}}
\def\midrule{\noalign{\medskip\hrule }}
\def\botrule{\noalign{\medskip\hrule }}
\setlength{\parskip}{\medskipamount}
 

\newcommand{\ee}{{\mathrm e}^+ {\mathrm e}^-}
\newcommand{\sq}{\tilde{\mathrm q}}
\newcommand{\seff}{\tilde{\mathrm f}}
\newcommand{\sele}{\tilde{\mathrm e}}
\newcommand{\sell}{\tilde{\ell}}
\newcommand{\snu}{\tilde{\nu}}
\newcommand{\ch}{\tilde{\chi}^\pm}
\newcommand{\chp}{\tilde{\chi}_{1}^+}
\newcommand{\chm}{\tilde{\chi}_{1}^-}
\newcommand{\chpm}{\tilde{\chi}_{1}^\pm}
\newcommand{\nt}{\tilde{\chi}^0}
\newcommand{\qq}{{\mathrm q}\bar{\mathrm q}}
\newcommand{\qqx}{{\mathrm q}\bar{\mathrm q'}}
\newcommand{\nunu}{\nu \bar{\nu}}
\newcommand{\mumu}{\mu^+ \mu^-}
\newcommand{\tautau}{\tau^+ \tau^-}
\newcommand{\ellell}{\ell^+ \ell^-}
\newcommand {\wenu} {{\mathrm{We}} \nu}
\newcommand{\nulqq}{\nu \ell {\mathrm q} \bar{\mathrm q}'}
\newcommand{\MZ}{M_{\mathrm Z}}
\newcommand{\MW}{M_{\mathrm W}}

\newcommand {\stopm}         {\tilde{\mathrm{t}}_{1}}
\newcommand {\stops}         {\tilde{\mathrm{t}}_{2}}
\newcommand {\stopbar}       {\bar{\tilde{\mathrm{t}}}_{1}}
\newcommand {\stopx}         {\tilde{\mathrm{t}}}
\newcommand {\sneutrino}     {\tilde{\nu}}
\newcommand {\slepton}       {\tilde{\ell}}
\newcommand {\stopl}         {\tilde{\mathrm{t}}_{\mathrm L}}
\newcommand {\stopr}         {\tilde{\mathrm{t}}_{\mathrm R}}
\newcommand {\stoppair}      {\tilde{\mathrm{t}}_{1} \bar{\tilde{\mathrm{t}}}_{1}}
\newcommand {\gluino}        {\tilde{\mathrm g}}

\newcommand {\sbotm}         {\tilde{\mathrm{b}}_{1}}
\newcommand {\sbots}         {\tilde{\mathrm{b}}_{2}}
\newcommand {\sbotbar}       {\bar{\tilde{\mathrm{b}}}_{1}}
\newcommand {\sbotx}         {\tilde{\mathrm{b}}}
\newcommand {\sbotl}         {\tilde{\mathrm{b}}_{\mathrm L}}
\newcommand {\sbotr}         {\tilde{\mathrm{b}}_{\mathrm R}}
\newcommand {\sbotpair}      {\tilde{\mathrm{b}}_{1} \bar{\tilde{\mathrm{b}}}_{1}}

\newcommand {\neutralino}    {\tilde{\chi }^{0}_{1}}
\newcommand {\neutrala}      {\tilde{\chi }^{0}_{2}}
\newcommand {\neutralb}      {\tilde{\chi }^{0}_{3}}
\newcommand {\neutralc}      {\tilde{\chi }^{0}_{4}}
\newcommand {\bino}          {\tilde{\mathrm B}^{0}}
\newcommand {\wino}          {\tilde{\mathrm W}^{0}}
\newcommand {\higginoa}      {\tilde{\mathrm H_{1}}^{0}}
\newcommand {\higginob}      {\tilde{\mathrm H_{1}}^{0}}
\newcommand {\chargino}      {\tilde{\chi }^{\pm}_{1}}
\newcommand {\charginop}     {\tilde{\chi }^{+}_{1}}
\newcommand {\KK}            {{\mathrm K}^{0}-\bar{\mathrm K}^{0}}
\newcommand {\ff}            {{\mathrm f} \bar{\mathrm f}}
\newcommand {\bq}            {\mathrm b} 
\newcommand {\cq}            {\mathrm c} 
\newcommand {\ele}           {\mathrm e} 
\newcommand {\bstopm} {\mbox{$\boldmath {\tilde{\mathrm{t}}_{1}} $}}
\newcommand {\Mt}            {m_{\mathrm t}}
\newcommand {\mb}            {m_{\mathrm b}}
\newcommand {\mc}            {m_{\mathrm c}}
\newcommand {\mscalar}       {m_{0}}
\newcommand {\Mgaugino}      {M_{1/2}}
\newcommand {\tanb}          {\tan \beta}
\newcommand {\rs}            {\sqrt{s}}
\newcommand {\WW}            {{\mathrm W}^+{\mathrm W}^-}
\newcommand {\eetautau}      {\ee-\rightarrow {\tau^+}{\tau^-}}
\newcommand {\MGUT}          {M_{\mathrm{GUT}}}
\newcommand {\Zboson}        {{\mathrm Z}^{0}}
\newcommand {\Wpm}           {{\mathrm W}^{\pm}}
\newcommand {\Wp}            {{\mathrm W}^{+}}
\newcommand {\allqq}         {\sum_{q \neq t} q \bar{q}}
\newcommand {\mixstop}       {\theta _{\stopx}}
\newcommand {\mixsbot}       {\theta _{\sbotx}}
\newcommand {\phiacop}       {\phi _{\mathrm {acop}}}
\newcommand {\cosjet}        {\cos\thejet}
\newcommand {\costhr}        {\cos\thethr}
\newcommand {\djoin}         {d_{\mathrm{join}}}
\newcommand {\mchar}         {m_{\chpm}}
\newcommand {\mstop}         {m_{\stopm}}
\newcommand {\msbot}         {m_{\sbotm}}
\newcommand {\mchi}          {m_{\neutralino}}
\newcommand {\pp}{p \bar{p}}
 
\newcommand{\gsim}{\;\raisebox{-0.9ex}
           {$\textstyle\stackrel{\textstyle >}{\sim}$}\;}
\newcommand{\lsim}{\;\raisebox{-0.9ex}{$\textstyle\stackrel{\textstyle<}
           {\sim}$}\;}

\newcommand{\degree}    {^\circ}
%
\newcommand{\Ecm}       {E_{\mathrm{cm}}}
\newcommand{\Ebeam}     {E_{\mathrm{b}}}
\newcommand{\roots}     {\sqrt{s}}
%
%
\newcommand{\thrust}    {T}
\newcommand{\nthrust}   {\hat{n}_{\mathrm{thrust}}}
\newcommand{\thethr}    {\theta_{\,\mathrm{thrust}}}
\newcommand{\phithr}    {\phi_{\mathrm{thrust}}}
\newcommand{\acosthr}   {|\cos\thethr|}
\newcommand{\thejet}    {\theta_{\,\mathrm{jet}}}
\newcommand{\acosjet}   {|\cos\thejet|}
\newcommand{\thmiss}    { \theta_{\mathrm{miss}} }
\newcommand{\cosmiss}   {| \cos \thmiss |}
\newcommand{\pbinv}     {\mathrm{pb}^{-1}}
%
%
\newcommand{\Evis}      {E_{\mathrm{vis}}}
\newcommand{\Rvis}      {E_{\mathrm{vis}}\,/\roots}
\newcommand{\Mvis}      {M_{\mathrm{vis}}}
\newcommand{\Rbal}      {R_{\mathrm{bal}}}
\newcommand{\mjet}      {\bar{M}_{\mathrm{jet}}}
%
%
%
\newcommand{\PhysLett}  {Phys.~Lett.}
\newcommand{\PRL} {Phys.~Rev.\ Lett.}
\newcommand{\PhysRep}   {Phys.~Rep.}
\newcommand{\PhysRev}   {Phys.~Rev.}
\newcommand{\NPhys}  {Nucl.~Phys.}
\newcommand{\NIM} {Nucl.~Instr.\ Meth.}
\newcommand{\CPC} {Comp.~Phys.\ Comm.}
\newcommand{\ZPhys}  {Z.~Phys.}
\newcommand{\IEEENS} {IEEE Trans.\ Nucl.~Sci.}
%
%
\newcommand{\OPALColl}  {OPAL Collab.}
\newcommand{\JADEColl}  {JADE Collab.}
\newcommand{\etal}      {{\it et~al.}}
\newcommand{\onecol}[2] {\multicolumn{1}{#1}{#2}}
\newcommand{\ra}        {\rightarrow}   
 
 
 
 
\begin{titlepage}
%
%
\begin{center}
    \large
    EUROPEAN LABORATORY FOR PARTICLE PHYSICS
\end{center}
\begin{flushright}
    \large
    \PPEnum\\
    \Date
\end{flushright}
%
%
%
%
\begin{center}
    \huge\bf\boldmath
    Search for Scalar Top \\
    and Scalar Bottom Quarks \\
    at $\rs$ = 189~GeV at LEP
\end{center}
\bigskip
\bigskip
\bigskip
%
%
\begin{center}
    \LARGE
    The OPAL Collaboration \\
\end{center}
\bigskip
\bigskip
\bigskip
%
%
\begin{abstract}
Searches for a scalar top quark and a scalar bottom quark 
have been performed using a data sample of 182~pb$^{-1}$
at a centre-of-mass energy of $\roots = $189~GeV
collected with the OPAL detector at LEP.
No evidence for a signal was found.
The 95\% confidence level (C.L.) lower limit
on the scalar top quark mass is  
90.3~GeV if the mixing angle between the supersymmetric partners 
of the left- and right-handed states of the top quark is zero.
In the worst case, when the scalar top quark decouples 
from the $\Zboson$ boson, the lower limit is 87.2 GeV.
These limits were obtained assuming that
the scalar top quark decays into 
a charm quark and the lightest neutralino,
and that the mass difference between 
the scalar top quark and the lightest 
neutralino is larger than 10~GeV.
The complementary decay mode of the scalar top quark 
decaying into a bottom quark, 
a charged lepton and a scalar neutrino has also been studied. 
From a search for the scalar bottom quark, 
a mass limit of 88.6~GeV was obtained if the mass difference
between the scalar bottom quark and the lightest 
neutralino is larger than 7~GeV\@.
These limits significantly improve the previous OPAL limits.
\end{abstract}
 
\bigskip
\bigskip
\begin{center}
{\large (\Submit) }\\
\bigskip
\end{center}
 
\bigskip
 
\end{titlepage}

\begin{center}{\Large        The OPAL Collaboration
}\end{center}\bigskip
\begin{center}{
G.\thinspace Abbiendi$^{  2}$,
K.\thinspace Ackerstaff$^{  8}$,
G.\thinspace Alexander$^{ 23}$,
J.\thinspace Allison$^{ 16}$,
N.\thinspace Altekamp$^{  5}$,
K.J.\thinspace Anderson$^{  9}$,
S.\thinspace Anderson$^{ 12}$,
S.\thinspace Arcelli$^{ 17}$,
S.\thinspace Asai$^{ 24}$,
S.F.\thinspace Ashby$^{  1}$
% ...
}\end{center}\bigskip
\bigskip
$^{  1}$School of Physics and Astronomy, University of Birmingham,
Birmingham B15 2TT, UK
\newline
$^{  2}$Dipartimento di Fisica dell' Universit\`a di Bologna and INFN,
I-40126 Bologna, Italy
\newline
% ...
$^{  5}$Cavendish Laboratory, Cambridge CB3 0HE, UK
\newline
% ...
$^{  8}$CERN, European Organisation for Particle Physics,
CH-1211 Geneva 23, Switzerland
\newline
$^{  9}$Enrico Fermi Institute and Department of Physics,
University of Chicago, Chicago IL 60637, USA
\newline
% ...
$^{ 12}$Indiana University, Department of Physics,
Swain Hall West 117, Bloomington IN 47405, USA
\newline
% ...
$^{ 16}$Department of Physics, Schuster Laboratory, The University,
Manchester M13 9PL, UK
\newline
$^{ 17}$Department of Physics, University of Maryland,
College Park, MD 20742, USA
\newline
% ...
$^{ 23}$Department of Physics and Astronomy, Tel Aviv University,
Tel Aviv 69978, Israel
\newline
$^{ 24}$International Centre for Elementary Particle Physics and
Department of Physics, University of Tokyo, Tokyo 113-0033, and
Kobe University, Kobe 657-8501, Japan
\newline
% ...
\newline

\section{Introduction}

Supersymmetric (SUSY) extensions of the Standard Model
predict the existence of bosonic partners of all known fermions.
The scalar top quark~($\stopx$), which is the bosonic partner of the 
top quark,
may be light
because of supersymmetric radiative corrections~\cite{stop1}\@. 
Furthermore, the supersymmetric partners of the
right-handed and left-handed top quarks
($\stopr$ and $\stopl$) mix, and
the resultant two mass eigenstates ($\stopm$ and $\stops$)
have a mass splitting, which 
may be very large due to the large top quark mass.  
Then the lighter mass eigenstate ($\stopm$),
$\stopm = \stopl \cos \mixstop + \stopr \sin \mixstop$,
where $\mixstop$ is a mixing angle,
can be lighter than any other
charged SUSY particle, and also lighter than the top quark~\cite{stop1}\@.
All SUSY breaking parameters are hidden 
in the $\mixstop$ and the mass of $\stopm$\@.

The scalar bottom quark ($\sbotx$) 
can also be light if $\tanb$, the ratio of 
vacuum expectation values of the two Higgs doublet fields, is larger than 
approximately 40.
In this case, the analogous mixing between the supersymmetric partners 
of the right- and left-handed states 
of the bottom quark ($\sbotr$ and $\sbotl$) becomes large,
and the resultant two mass eigenstates ($\sbotm$ and $\sbots$)
also have a large mass splitting~\cite{bartl}\@.
The mass of the lighter mass eigenstate 
($\sbotm$) may therefore be within the reach of LEP2. 

Assuming R-parity~\cite{RP} conservation,
the dominant decay mode of the $\stopm$ is  
expected to be either
$\stopm \ra \cq \neutralino$ or $\stopm \ra \bq \snu \ell^{+}$,
where $\neutralino$ is 
the lightest neutralino and $\snu$ is the scalar neutrino.
The latter decay mode
is dominant, if it is kinematically allowed.
Otherwise the flavour changing two-body decay, 
$ \stopm \ra \cq \neutralino$, is dominant
except for the small region of $m_{\stopm}-\mchi > m_{\Wpm} + \mb$\footnote{
In this region, $\stopm \ra \bq  \neutralino \Wp$(on shell) becomes
dominant through a virtual chargino as described in Section~4.
This decay mode has not been searched for in this paper.}\@.
Both of these decay modes ($\stopm \ra \cq \neutralino$
and $\stopm \ra \bq \snu \ell^{+}$) have been searched for. 
The dominant decay mode of the $\sbotm$ is 
expected to be $ \sbotm \ra \bq \neutralino$\@.
Under the assumption of  R-parity conservation,
$\neutralino$ and $\snu$ are invisible in the detector.
Thus, $\stoppair$ and $\sbotpair$ events 
are characterised by two acoplanar 
jets\footnote{Two jets are called `acoplanar' if they
not back-to-back with each other in the plane 
perpendicular to the beam axis.}
or two acoplanar jets plus
two leptons, with missing energy.
The phenomenology of the production and decay of 
$\stopm$ ($\sbotm$) is described in Section 2 of 
Ref.~\cite{stop171}\@.

The D0 Collaboration has reported a lower limit~\cite{d0} 
on the $\stopm$ mass of about 85~GeV (95\% C.L.)
for the case that $\stopm \ra \cq \neutralino$ is the dominant decay mode
and the mass difference 
between $\stopm$ and $\neutralino$ is larger than about 35~GeV\@.
Searches at $\ee$ colliders are sensitive to a smaller mass difference. 
Mass limits for the $\stopm$ have already been obtained around 
the $\Zboson$ peak (LEP1) assuming 
$\stopm \ra \cq \neutralino$~\cite{opalstop}\@.
A 95\% C.L. lower limit of 76~GeV
for a mass difference larger than 5~GeV 
has been obtained as a result 
of previous searches at centre-of-mass energies 
161~\cite{stop161}, 171~\cite{stop171,alephstop} and
183~GeV~\cite{stop183,ALEPH183}\@.

In 1998 the LEP $\ee$ collider at CERN was operated 
at $\rs$= 188.6~GeV,
and a data sample of 182.1~$\pbinv$ was collected with
the OPAL detector. 
In this paper direct searches for $\stopm$ and $\sbotm$
using this data sample are reported.
The results shown here have been obtained by combining the results
obtained at this new centre-of-mass energy 
with those previously obtained by the OPAL detector at $\rs$ 
= 161, 171 and 183~GeV \cite{stop161,stop171,stop183}\@. 

\section{The OPAL Detector and Event Simulation}

The OPAL detector,
which is described in detail in Ref.~\cite{OPAL-detector},
is a multipurpose apparatus
having nearly complete solid angle coverage.
The central detector consists of
a silicon strip detector and tracking chambers,
providing charged particle tracking 
for over 96\% of the full solid 
angle,
inside a uniform solenoidal magnetic field of 0.435~T\@.
A lead-glass electromagnetic calorimeter (EM)  
located outside the magnet coil
is hermetic
in the polar angle\footnote{A right-handed 
coordinate system is adopted,
where the $x$-axis points to the centre of the LEP ring,
and positive $z$ is along  the electron beam direction.
The angles $\theta$ and $\phi$ are the polar and azimuthal angles,
respectively.} range of $|\cos \theta |<0.82$ for the barrel
region and $0.81<|\cos \theta |<0.984$ for the endcap region.
The magnet return yoke
consisting of barrel and endcap sections along with
pole tips 
is instrumented for hadron calorimetry (HCAL)
in the region $|\cos \theta |<0.99$\@.
Four layers of muon chambers cover the outside of the hadron calorimeter.
Calorimeters close to the beam axis measure the luminosity
using small angle Bhabha scattering events
and complete the geometrical acceptance down to 24 mrad.

Monte Carlo simulation of the production and decay
of $\stopm$($\sbotm$)
was performed as follows~\cite{stopgen}\@.
The $\stoppair$($\sbotpair$) pairs were generated
taking into account initial-state radiation~\cite{lund}\@.
The hadronisation process was subsequently performed
to produce colourless $\stopm$-hadrons ($\sbotm$-hadrons)
and other fragmentation products
according to the Lund string fragmentation scheme
(JETSET 7.4)~\cite{lund,fragment}\@.
The parameters for perturbative QCD and fragmentation processes
were optimised using the hadronic $\Zboson$ decays
measured by OPAL~\cite{opalfragment}\@.
For the fragmentation of $\stopm$($\sbotm$),
the fragmentation function proposed
by Peterson {\it et al.}~\cite{lund,Peterson} was used.
The $\stopm$-hadron ($\sbotm$-hadron) was formed from a $\stopm$-quark 
($\sbotm$-quark) and a spectator anti-quark or diquark.
For the $\stopm$($\sbotm$) decaying into
$\cq \neutralino$ ($\bq \neutralino$),
a colour string was stretched between the charm quark (the bottom quark)
and the spectator.
This colour singlet system 
was hadronised using the Lund scheme~\cite{lund,fragment}\@.
Gluon bremsstrahlung was allowed
in this process, and 
the Peterson function was also used 
for the charm quark and the bottom quark fragmentation.
The signals for the decays $ \stopm \ra \bq \ell^{+} \snu $ 
were simulated in a similar manner.

One thousand events were generated at each point of a two dimensional grid
of spacing of generally 5~GeV step in $(m_{\stopm}, m_{\neutralino})$ 
for $\stopm \ra \cq \neutralino$, 
in $(m_{\stopm}, m_{\snu})$ for $\stopm \ra \bq \ell^{+} \snu$ and 
$\stopm \ra \bq \tau^{+} \snu$,
and $(m_{\sbotm}, m_{\neutralino})$ for $\sbotm \ra \bq \neutralino$\@.
Smaller steps were used for the case of small mass differences
($\Delta m  =  \mstop - \mchi$, $ \mstop - m_{\snu}$ or
$ m_{\sbotm} - m_{\neutralino}$)\@. 
The mixing angles of the $\stopm$ and $\sbotm$ were set to
zero when these events were generated.
The dependence of the detection efficiencies 
on these mixing angles is taken into
account as a systematic error as described in Ref.~\cite{stop183}\@.

The background processes were simulated as follows.
The PYTHIA~\cite{lund} generator
was used to simulate multihadronic ($\qq(\gamma)$) events,
and KORALZ~\cite{KORALZ} to generate $\tau^+ \tau^- (\gamma)$
and $\mumu (\gamma)$ events.  
Bhabha events, $\ee \ra \ee (\gamma) $, were generated with 
the BHWIDE program~\cite{BHWIDE}\@.
Two-photon processes are the most important background
for the case of small mass differences,
since in such cases signal events have small visible energy and 
small transverse momentum relative to
the beam direction.
Using the Monte Carlo generators 
PHOJET~\cite{PHOJET}, PYTHIA~\cite{lund} and HERWIG~\cite{HERWIG},
hadronic events from two-photon processes were simulated
in which the invariant mass of the photon-photon
system ($M_{\gamma \gamma}$) was larger than 5.0~GeV\@.
Monte Carlo samples for four-lepton events ($\ee \ee$, $\ee \mumu$ and 
$\ee \tautau$) were generated with the Vermaseren 
program~\cite{Vermaseren}\@.
The grc4f generator~\cite{grace} was used for 
all four-fermion processes except for regions
covered by the two-photon simulations.
All interference effects of the various diagrams
are taken into account in grc4f.
Four-fermion processes in which at least one
of the fermions is a neutrino constitute a serious background
at large mass differences.
The dominant contributions come from  
${\mathrm W}^+ {\mathrm W}^-$, $\gamma ^{*} \Zboson$ and ${\mathrm W}e\nu$
events.
The Excalibur~\cite{excalibur} and PYTHIA~\cite{lund} generators
were also used to study uncertainties in the grc4f generator.
The generated signal and background events were processed
through the full simulation of the OPAL detector~\cite{GOPAL},
and the same analysis chain was applied as to the data.

\section{Analysis}
 
Since the event topologies of $\stopm \ra \cq \neutralino$ 
and $\sbotm \ra \bq \neutralino $ are similar, 
the same selection criteria were used (Section~3.1, analysis A)\@. 
In Section 3.2 (analysis B), the selection criteria for
$\stopm \ra \bq \ell^{+} \snu$ are discussed. 
These analyses are similar to those in Ref.~\cite{stop183},
and the quality criteria therein were used
to select good tracks and clusters.
Variables used for the cuts, such as 
the total visible energy and the total transverse momentum,
were calculated as follows.
First, the four-momenta of the tracks and 
those of the EM and HCAL clusters not
associated with charged tracks were summed.
Whenever a calorimeter cluster had associated charged tracks,
the expected energy deposited by the tracks was subtracted
from the cluster energy to reduce double counting.
If the energy of a cluster was smaller
than the expected energy deposited by the associated tracks,
the cluster energy was not used.

The following three preselections, 
which are common to analyses A and B,
were applied: 
{\bf(1)}
The number of charged tracks was required to be at least four. 
The ratio of the number of good tracks
to the total number of reconstructed tracks was required 
to be greater than 0.2 
to reduce beam-gas and beam-wall background events.
The visible mass of the event  
was also required to be larger than 3~GeV\@.
{\bf(2)}
To reduce the background from two-photon processes,
the energy deposited had to be less than 2~GeV in each silicon tungsten
forward calorimeter, less than 2~GeV in each forward detector 
and less than 5~GeV in each side of the gamma-catcher detector.
These detectors are located in the forward region
($|\cos \theta|>0.98$) surrounding the beam pipe.
{\bf(3)}
The visible energy in the region of $|\cos \theta|>0.9$
was required to be less than 10\% of the total visible energy.
In addition,
the polar angle of the missing momentum direction, $\thmiss$,
was required to satisfy $\cosmiss < 0.9$
to reduce the two-photon and the $\qq(\gamma)$ background.

\subsection{Analysis A: {\boldmath$\boldsymbol{\stopm \ra \cq \neutralino}$} 
and {\boldmath$\boldsymbol{\sbotm \ra \bq \neutralino}$}} 
 
The experimental signature for 
$\stoppair$($\stopm \ra \cq \neutralino$) events
and $\sbotpair$ events
is an acoplanar two-jet topology with a large transverse 
momentum with respect to the beam axis.
The fragmentation functions of $\stopm $ and $\sbotm $ are expected to be
hard and the invariant mass of the charm (or bottom) quark and
the spectator quark is small,
therefore the jets are expected to be narrow.
 
The following five selections were applied:
{\bf(A1)}
Events from two-photon
processes were largely removed by demanding that the event transverse
momentum with respect to the beam axis, $P_t$, be greater than 4.5~GeV\@.
Since the hadron calorimeter, with its limited energy resolution, gives
fluctuations in energy measurement, this selection
was applied to $P_t$ calculated both with and without
the hadron calorimeter.
Fig.~1(a) shows the distribution of
$P_t$ calculated with the hadron calorimeter after the preselections.
{\bf(A2)}
The number of reconstructed jets was required to be exactly two.
Jets were reconstructed using the Durham algorithm~\cite{DURHAM} 
with the jet resolution parameter of 
$y_{\rm cut}$ = $0.005 (\Evis / \rs)^{-1}$,
where $\Evis$ is the total visible energy.
This $\Evis$-dependent $y_{\rm cut}$ parameter
was necessary for good jet reconstruction over a wide range of 
$\mstop$, $\msbot$ and $\mchi$\@.
Furthermore, both reconstructed jets were required to contain 
at least two charged particles
to reduce the $\tau^{+} \tau^{-}$ background where
at least one of the $\tau$'s decayed into only one charged particle.
{\bf(A3)}
The acoplanarity angle, $\phiacop$,
is defined as $\pi$ minus the azimuthal opening angle
between the directions of the two reconstructed jets.
To ensure the reliability of the calculation of $\phiacop$,
both jet axes were required to have a polar angle satisfying
$|\cos{\theta}_{\rm jet}| < 0.95$\@.
The value of $\phiacop$ was required to be larger than 20$\degree$\@.
{\bf(A4)}
`Softness' was defined as 
($\frac{{M}_1}{{E}_1} + \frac{{M}_2}{{E}_2}$),
where ${M}_1$ and ${M}_2$ are the invariant masses 
of the two reconstructed jets, 
and ${E}_1$ and ${E}_2$ are the energies of the jets.
The signal events have low values of `Softness',  
whereas two-photon events
which pass the acoplanarity cut have relatively large values
as shown in Fig.~4 in Ref.~\cite{stop183}\@. 
It was required that $1.5 \times {\rm Softness} < (P_t-4.5)$,
where $P_t$ is calculated with the hadron calorimeter and 
given in units of GeV\@.
{\bf(A5)}
The arithmetic mean of the invariant masses of the jets, $\mjet$, 
was required to be smaller than 8~GeV\@.
When the invariant mass of the event, $\Mvis$, was larger than 65~GeV,
a harder cut $\mjet < $ 5~GeV was applied to 
reduce background from $\wenu$ events.
Fig.~1(b) shows the $\mjet$ distributions for 
data, the simulated background processes and
typical $\stoppair$ events.
As shown in this figure,
jets from $\stopm$ are expected to have
low invariant masses.

The numbers of events remaining after each cut are listed
in Table~\ref{tab:nevA}\@.
The table also shows 
the corresponding numbers of simulated events
for background processes and 
for two samples of simulated $\stoppair$ 
($\stopm \ra \cq \neutralino$) and one sample of $\sbotpair$ events.

\begin{table}[h]
\centering
\begin{tabular}{|l||r||r||r|r|r|r||r|r|r|}
\hline
   &  \multicolumn{1}{c||}{data} & \multicolumn{1}{c||}{total} &
$\qq (\gamma)$ & $\ellell (\gamma)$ & 
\multicolumn{1}{|c|}{`$\gamma \gamma$'} &
\multicolumn{1}{c||}{4-f} &
\multicolumn{3}{c|}{\rule{0mm}{6mm} $\stoppair$ and $\sbotpair$}  \\
    &    &   \multicolumn{1}{c||}{bkg.} 
&  &  &  &  & \multicolumn{3}{c|}{ }   \\
\hline
$m_{\stopm}$ (GeV)&       &      &         &          &            &
        & \multicolumn{1}{c|}{90} & 
          \multicolumn{1}{c|}{90} &
          \multicolumn{1}{c|}{--}  \\
$m_{\sbotm}$ (GeV)&       &      &         &          &            &
        & \multicolumn{1}{c|}{--} & 
          \multicolumn{1}{c|}{--} &
          \multicolumn{1}{c|}{90}  \\
$m_{\neutralino}$ (GeV)&       &      &         &            &            &
        & \multicolumn{1}{c|}{85} & 
          \multicolumn{1}{c|}{70} & 
          \multicolumn{1}{c|}{70}  \\
\hline
 cut (A1)
& 4073 & 4274 & 2157 &  497 &  121 & 1499 & 397 & 685 & 707 \\
\hline
 cut (A2)
&  995 & 1048 &  857 & 33.2 & 37.4 &  119 & 239 & 611 & 668 \\
\hline
 cut (A3)
&   75 & 83.7 & 0.18 & 0.25 & 7.7 & 75.6 & 237 & 564 & 609 \\
\hline
 cut (A4)
&   75 & 78.1 & 0.18 & 0.25 & 2.1 & 75.6 & 176 & 564 & 606 \\
\hline
 cut (A5)
&    4 & 6.9 & 0.00 & 0.09 & 1.5 & 5.3 & 176 & 560 & 595 \\
& & ($\pm 1.0) $ & ($~^{+0.04}_{-0.00}$) & ($\pm 0.04) $ & ($\pm 0.9$) & 
($\pm 0.4$) & & & \\
\hline
\end{tabular}
\caption[]
{
Numbers of events remaining after each cut 
for various background processes are compared with data.
The simulated background processes were normalised
to the integrated luminosity of the data.
The errors due to Monte Carlo statistics are also shown.  
Numbers for 
three simulated event samples of $\stoppair$ and $\sbotpair$ are 
also given (each starting from 1000 events).
}
\label{tab:nevA}
\end{table}
 
After all cuts, four events were observed in the data,
which is consistent with the expected number of background events
of 6.9$\pm$1.0, mainly from four-fermion processes.

The efficiencies for both $\stoppair$ and $\sbotpair $ events
are 30--60\%,
if the mass difference between $\stopm$($\sbotm$) 
and $\neutralino $ is larger than 10~GeV\@.
A modest efficiency of about 20\% is obtained for 
a mass difference of 5~GeV for $\stoppair$ events.
An additional efficiency loss of 3\% (relative) arose from beam-related
background in the silicon tungsten forward calorimeter, 
forward detector and gamma-catcher detectors,
which was estimated using random beam crossing events.
This inefficiency was taken into account
in the limit calculation.

\subsection{Analysis B: {\boldmath$\boldsymbol{\stopm \ra \bq \ell \snu}$}} 
 
The experimental signature for $\stoppair$($\stopm \ra \bq \ell \snu$) events
is two acoplanar jets plus two leptons with missing transverse 
momentum with respect to the beam axis.
The momenta of the leptons and the missing transverse momentum
depend strongly on the mass difference
between $\stopm$ and $\snu$\@.
To obtain optimal performance, 
two sets of selection criteria (analyses B-L and B-H) 
were applied for small and large mass differences, respectively.

The numbers of events remaining after each cut are listed
in Tables~\ref{tab:nevBL} and \ref{tab:nevBH}\@.
The tables also show 
the corresponding numbers for the simulated background processes
and for the simulated $\stoppair$ signals.

\subsubsection{Small mass difference case} 

For the case of a small mass difference ($\Delta m \leq$ 10~GeV), 
the following four selection criteria were applied:
{\bf(B-L1)}
$P_t$ was required to be greater than 5~GeV\@.
{\bf(B-L2)} 
The number of charged tracks was required to be at least six.
Furthermore,
the number of reconstructed jets was required to be at least four,
since the signal would contain two hadronic jets 
plus two isolated leptons.
Jets were reconstructed using the Durham algorithm~\cite{DURHAM} with 
the jet resolution parameter $y_{\mathrm{cut}}$ = 0.004\@.
{\bf(B-L3)} 
To examine the acoplanarity of the events remaining, jets were 
reconstructed using the Durham algorithm
where the number of jets was forced to be two.
To ensure a good measurement of the acoplanarity angle,
$|\cos{\theta}_{\rm jet}| < 0.95$ was required for 
both reconstructed jets.
Finally, the acoplanarity angle, $\phiacop$, between these two jets
was required to be greater than 15$\degree$\@.
In the three-body decay, the transverse momentum carried by the $\snu$ 
with respect to the original $\stopm$-momentum 
is smaller than that of $\neutralino$ in the two-body decay.
When the $\stopm$ is light, the outgoing $\snu$ is strongly boosted 
toward the direction of the parent $\stopm$
and $\phiacop$ for the signal becomes small.
This is the reason why a looser acoplanarity angle cut was used.
Fig.~1(c) shows the $\phiacop$ distributions for 
the data, the simulated background processes and
typical $\stoppair$ events.
{\bf(B-L4)} 
The total visible energy, $\Evis$,
was required to be smaller than 60~GeV to reject
four-fermion events.

\begin{table}[h]
\centering
\begin{tabular}{|l||r||r||r|r|r|r||r|r|}
\hline
   &  \multicolumn{1}{c||}{data}         & \multicolumn{1}{c||}{total} 
   &  \multicolumn{1}{c|}{$\qq (\gamma)$} 
   &  \multicolumn{1}{c|}{$\ellell (\gamma)$} 
   &  \multicolumn{1}{c|}{`$\gamma \gamma$'} 
   &  \multicolumn{1}{c||}{4-f} 
   &  \multicolumn{2}{c|}{\rule{0mm}{6mm} $\stoppair$}  \\
   &       & \multicolumn{1}{c||}{bkg.} 
   &  &  &  &  & \multicolumn{2}{c|}{ }   \\
\hline 
$m_{\stopm}$ (GeV)&       &      &         &          &            &
        & \multicolumn{1}{c|}{75} & 
          \multicolumn{1}{c|}{90} \\
$m_{\snu}$ (GeV)&       &      &         &            &            &
        & \multicolumn{1}{c|}{68} & 
          \multicolumn{1}{c|}{80} \\
\hline 
 cut (B-L1)
& 3900 & 4058 & 2030 &  480 & 88.7 & 1460 & 153 & 413 \\
\hline
 cut (B-L2)
& 1016 & 1023 &  307 & 0.11 & 6.4 &  709 & 132 & 373 \\
\hline
 cut (B-L3)
&  216 &  217 & 10.5 & 0.02 & 1.7 &  205 & 99 & 339 \\
\hline
 cut (B-L4)
&    0 & 2.1 & 0.04 & 0.00 & 1.7 & 0.4 & 99 & 339 \\
& & ($\pm 0.9) $ & ($\pm 0.04) $ & ($~^{+0.02}_{-0.00}$) & ($\pm 0.9$) & 
($\pm 0.1$) & &  \\
\hline
\end{tabular}
\caption[]{
Numbers of events remaining after each cut 
for various background processes are compared with data.
The simulated background processes
were normalised to the integrated luminosity of the data.
The errors due to Monte Carlo statistics are also shown.  
Numbers for
two simulated samples of $\stoppair$ are 
also given (each starting from 1000 events).
In these samples, the branching fraction to each lepton flavour
is assumed to be the same.
}
\label{tab:nevBL}
\end{table}

No events were observed in the data after all the cuts. 
This is consistent with the number of expected background
events (2.1$\pm$0.9), mainly from two-photon background.
The detection efficiencies are 30-35\%
if the mass difference between $\stopm$ and $\snu$
is 10~GeV, 
and if the branching fraction to each lepton flavour
is the same.
Even if the
branching fraction into $\bq \tau ^{+} \snu_{\tau} $
is 100\%, the efficiencies are 25-30\%.

\subsubsection{Large mass difference case} 

The selection criteria for a large mass difference ($\Delta m > $ 10~GeV) 
were as follows:
{\bf(B-H1)}
$P_t$ was required to be greater than 6~GeV\@.
{\bf(B-H2)} 
The number of charged tracks was required to be at least six, 
and the number of reconstructed jets was required 
to be at least three.
Jets were reconstructed with 
the same jet resolution parameter as in (B-L2)\@.
{\bf(B-H3)} 
The same selection as (B-L3) was applied on the $\phiacop$ variable
to reject $\qq(\gamma)$ events. 
{\bf(B-H4)} 
A candidate event was required to contain at least one lepton,
since a signal event would contain two isolated leptons.
The selection criteria for leptons are given
in Ref.~\cite{stop183}\@.
{\bf(B-H5)} 
The invariant mass of the event excluding the most energetic lepton,
$M_{\mathrm{hadron}}$,
was required to be smaller than 60~GeV
in order to reject $\WW \ra \nulqq $ events. 
As shown in Fig.~1(d), a large fraction of four-fermion
events was rejected using this requirement.
Furthermore the invariant mass excluding all identified leptons
was required to be smaller than 40~GeV\@.
{\bf(B-H6)} 
Finally, the visible mass of event, $\Mvis$, must be smaller than 80~GeV 
to reduce $\WW$ background events in which 
one of W decays into $\tau \nu$ and the other 
into ${\rm q \bar{q}^{'}(g)}$\@.
If one jet from ${\rm q \bar{q}^{'}(g)}$ was misidentified as a tau lepton,
this event could pass the previous cut (B-H5).

\begin{table}[h]
\centering
\begin{tabular}{|l||r||r||r|r|r|r||r|r|r|}
\hline
   &  \multicolumn{1}{c||}{data}         & \multicolumn{1}{c||}{total} 
   &  \multicolumn{1}{c|}{$\qq (\gamma)$} 
   &  \multicolumn{1}{c|}{$\ellell (\gamma)$} 
   &  \multicolumn{1}{c|}{`$\gamma \gamma$'} 
   &  \multicolumn{1}{c||}{4-f} 
   &  \multicolumn{3}{c|}{\rule{0mm}{6mm} $\stoppair$}  \\
   &       & \multicolumn{1}{c||}{bkg.} 
   &  &  &  &  & \multicolumn{3}{c|}{ }   \\
\hline 
$m_{\stopm}$ (GeV)&       &      &         &          &            &
        & \multicolumn{1}{c|}{90} & 
          \multicolumn{1}{c|}{90} &
          \multicolumn{1}{c|}{90}  \\
$m_{\snu}$ (GeV)&       &      &         &            &            &
        & \multicolumn{1}{c|}{80} & 
          \multicolumn{1}{c|}{70} & 
          \multicolumn{1}{c|}{45}  \\
\hline 
 cut (B-H1)
& 3576 & 3683 & 1802 &  448 & 53.2 & 1380 & 299 & 608 & 657 \\
\hline
 cut (B-H2)
& 2261 & 2399 & 1125 & 3.38 & 10.5 & 1261 & 299 & 605 & 647 \\
\hline
 cut (B-H3)
&  618 &  638 & 37.5 & 0.65 & 0.85 &  599 & 280 & 570 & 569 \\
\hline
 cut (B-H4)
&  444 &  478 & 15.5 & 0.53 & 0.00 &  462 & 239 & 534 & 541 \\
\hline
 cut (B-H5)
&    5 & 4.3 & 0.10 & 0.09 & 0.00 & 4.1 & 239 & 534 & 447 \\
\hline
 cut (B-H6)
&    3 & 1.9 & 0.10 & 0.04 & 0.00 & 1.8 & 239 & 534 & 437 \\
& & ($~^{+0.5}_{-0.3}$) & ($\pm 0.03) $ & ($\pm 0.03) $ & 
($~^{+0.5}_{-0.00}$) & 
($\pm 0.3$) & & & \\
\hline
\end{tabular}
\caption[]{
Numbers of events remaining after each cut
for various background processes are compared with data.
The simulated background processes were normalised
to the integrated luminosity of the data.
The errors due to Monte Carlo statistics are also shown.  
Numbers for
three simulated samples of $\stoppair$ are 
also given (each starting from 1000 events).
In these samples, the branching fraction to each lepton flavour
is assumed to be the same.
}
\label{tab:nevBH}
\end{table}
 
Three events were observed in the data,
which is consistent with the number of expected background
events~(1.9~$^{+0.5}_{-0.3}$)\@.
The dominant background arises from four-fermion processes.
The detection efficiencies are 25-60\%,
if the mass difference between the $\stopm$ 
and $\snu$ is 10~GeV, and if the $\snu$ is heavier than 30~GeV\@.
The detection efficiencies
for $\stopm \ra \bq \tau^{+} \snu_{\tau}$ were found to be slightly
smaller than in the case where the branching fraction to each lepton
flavour is assumed to be the same.
 
\section{Results}

The observed number of candidate events in each case is consistent with
the expected number of background events.
Since no evidence for $\stoppair$ and $\sbotpair$ pair-production 
has been observed,
lower limits on $\mstop$ and $\msbot$ are calculated.

The systematic errors on the expected number 
of signal and background events 
were estimated in the same manner as
in the previous paper~\cite{stop183}\@.
The main sources of systematic errors on signal 
are uncertainties in the $\stopm$ and $\sbotm$ fragmentation (5--10\%)
and in Fermi motion of the spectator quark (3--8\%)\@. 
The main sources of systematic errors on background 
are uncertainties in the generator of four-fermion processes (20\%) 
and statistical fluctuation in two-photon Monte Carlo samples.
Detailed descriptions are given in Ref.~\cite{stop183}\@.
Systematic errors are taken into account
when calculating limits~\cite{SYSTEM}\@.

Figure~2 shows the 95\% C.L. excluded regions
in the ($\mstop$ , $\mchi$) plane for $\stopm \ra \cq \neutralino$\@.
In this figure there is a triangular region of
$m_{\stopm}-\mchi > m_{\Wpm} + \mb$,
in which $\stopm \ra \bq  \neutralino \Wp$(on shell)
through a virtual chargino becomes dominant
even if the chargino is heavy.
This region is not excluded.
Since the momenta of $\bq$ and $\neutralino$ are small
at the current centre-of-mass energy,
the signal topology is very similar to $\WW$ background events. 

Figures~3(a) and (b) show the 95\% C.L. excluded regions
in the ($\mstop$ , $m_{\snu}$) plane for 
$\stopm \ra \bq \ell \snu$ ($\ell$= e,$\mu$,$\tau$)
and $\stopm \ra \bq \tau ^{+} \snu_{\tau} $, respectively.
The branching fraction to each lepton flavour $\ell^{+}$ depends
on the composition of the lightest chargino~\cite{stop171}.
As the chargino becomes Higgsino-like, the branching fraction into 
$\bq \tau ^{+} \snu_{\tau} $ becomes large.
In the limit that the chargino is a pure Wino state,
the branching fraction to each lepton flavour is the same.
Two extreme cases in which
the branching fraction to each lepton flavour is the same, or
the branching fraction into $\bq \tau ^{+} \snu_{\tau} $ is 100\%, 
were considered here\@.

The 95\% C.L. mass bounds of $\stopm$ are listed in Table~\ref{limit1}
for various values of $\mixstop$\@.
Assuming that $\stopm$ decays into $\cq \neutralino$,
and the mass difference between $\stopm$ and $\neutralino$
is greater than 10~GeV,
$\stopm$ is found to be heavier than 90.3~GeV for $\mixstop$ = 0.0\@. 
A lower limit of 87.2~GeV is obtained
even if $\stopm$ decouples from the $\Zboson$ boson ($\mixstop$=0.98~rad)\@. 
When $\stopm$ decays into $\bq \ell \snu$,
the lower limit on $\mstop$ is 90.5~GeV for the zero mixing angle case, 
assuming that the mass difference between $\stopm$ and $\snu$
is greater than 10~GeV
and that the branching fraction to each lepton flavour is the same.
These limits improve significantly (about 10~GeV)
the previous OPAL limits~\cite{stop183}\@.

\begin{table}[h] \centering
\begin{tabular}{|r || c | c || c | c | } \hline
\multicolumn{5}{|c|}{ \rule{0mm}{6mm} 
            Lower limit on $\mstop$ (GeV) } \\ \hline 
 & \multicolumn{2}{|c||}{ \rule{0mm}{6mm} $\stopm \ra \cq \neutralino$ }
 & \multicolumn{1}{|c|}{$\stopm \ra \bq \ell \snu$} 
 & \multicolumn{1}{|c|}{$\stopm \ra \bq \tau \snu_{\tau}$} \\ 
 & \multicolumn{2}{|c||}{ } 
 & \multicolumn{1}{|c|}{$\ell = {\mathrm e}, \mu, \tau$} 
 & \multicolumn{1}{|c|}{Br = 100\%} \\ \hline
 
$\mixstop$ (rad) & $ \Delta m \geq 5$~GeV & 
                 $ \Delta m \geq 10$~GeV  & $ \Delta m \geq 10$~GeV 
                 & $ \Delta m \geq 10$~GeV \\ \hline
0.0   &                           89.1 &  90.3 & 90.5 & 90.0  \\ \hline
$ \leq \frac{1}{8} \pi $ &        88.6 &  89.9 & 89.9 & 89.5  \\ \hline
$ \leq \frac{1}{4} \pi $ &        86.8 &  87.7 & 88.6 & 87.9  \\ \hline
 0.98 &                           86.4 &  87.2 & 88.0 & 87.5  \\ \hline
\end{tabular}
\caption[]{
The excluded $\mstop$ region at 95\% C.L. ($\Delta m  =  \mstop - \mchi$
 or $ \mstop - m_{\snu}$)\@. 
}
\label{limit1}
\end{table}

The 95\% C.L. excluded regions in the ($m_{\sbotm}$, $m_{\neutralino}$) 
plane are shown in Fig.~4 for two cases $\mixsbot$= 0 and 1.17~rad\@. 
The numerical mass bounds are listed in Table~\ref{limit2}
for various $\mixsbot$\@.
These bounds are significantly stronger than 
the previous OPAL limits~\cite{stop183} by about 10~GeV\@.
The lower limit on the $\sbotm$-mass is found to be 88.6~GeV,
if $\Delta m$ is greater than 7~GeV and $\mixsbot$ = 0.0\@.
When $\Delta m$ is greater than 10~GeV and $\neutralino$ is
heavier than 30~GeV, $\sbotm$ is found to be heavier than 89.8~GeV\@.
If the $\sbotm$ decouples from the $\Zboson$ boson ($\mixsbot$=1.17~rad), 
the lower limit is 74.9~GeV\@.
Since the electromagnetic charge of $\sbotm$ is half that of $\stopm$,
the coupling between $\gamma$ and $\sbotm$ is weaker
than between $\gamma$ and $\stopm$\@.
Therefore the production cross-section of $\sbotpair$ is strongly suppressed
when the $\sbotm$ decouples from the $\Zboson$ boson.

\begin{table}[h] \centering
\begin{tabular}{|r || c | c |} \hline
\multicolumn{3}{|c|}{ \rule{0mm}{6mm} Lower limit on $m_{\sbotm}$ (GeV)
($\sbotm \ra \bq \neutralino$) } \\ \hline
 
$\mixsbot$ (rad) & $ \Delta m \geq 7$~GeV &  $\Delta m \geq 10$~GeV \\ 
                 &                  & $ m_{\neutralino} \geq 30$~GeV \\  \hline
0.0              &               88.6      &      89.8              \\ \hline
$ \leq \frac{1}{8} \pi $ &       87.8      &      89.2              \\ \hline
$ \leq \frac{1}{4} \pi $ &       82.2      &      85.0              \\ \hline
1.17                     &       65.8      &      74.9              \\ \hline
\end{tabular}
\caption[]{
The excluded $\msbot$ region at 95\% C.L. 
($\Delta m  = m_{\sbotm} - m_{\neutralino}$)
} 
\label{limit2}
\end{table}
 
\section{Summary and Conclusion}
 
A data sample of 182.1~pb$^{-1}$ collected using the OPAL detector
at $\roots = $188.6~GeV has been analysed
to search for pair production of the scalar top quark and the 
scalar bottom quark predicted by supersymmetric theories
assuming R-parity conservation.
No evidence was found above the background level expected 
from the Standard Model.

The 95\% C.L. lower limit on the scalar top quark mass is 
90.3~GeV,  
if the mixing angle of the scalar top quark is zero\@.
If the $\stopm$ decouples from the $\Zboson$ boson, 
a lower limit of 87.2~GeV is obtained.
These limits were estimated assuming that
the scalar top quark decays into a charm quark and the lightest neutralino 
and that the mass difference between the scalar top and the lightest 
neutralino is larger than 10~GeV\@.

Assuming a relatively light scalar neutrino
($ m_{\snu} \leq \, \mstop - m_{\bq}$), 
the complementary decay mode, in which the scalar top quark decays  
into a bottom quark, a charged lepton 
and a scalar neutrino, has also been studied. 
If the mass difference between the scalar top quark and the scalar neutrino
is greater than 10~GeV
and if the mixing angle of the scalar top quark is zero,
the 95\% C.L. lower limit on the scalar top quark mass is 90.5~GeV\@.
This limit is obtained assuming that the branching fraction to
each lepton flavour is the same.
If the branching fraction to the tau lepton is 100\%,
a lower limit of 90.0~GeV is obtained. 

The 95\% C.L. mass limit on the light scalar bottom quark
is found to be 88.6~GeV,
assuming that the mass difference between the scalar bottom quark and 
the lightest neutralino is greater than 7~GeV
and that the mixing angle of the scalar bottom quark is zero.
If the mass difference is greater than 10~GeV
and the lightest neutralino is heavier than 30~GeV,
the mass limit on the light scalar bottom quark is 89.8~GeV
for zero mixing angle.
If the scalar bottom quark decouples from the $\Zboson$ boson,
a lower limit of 74.9~GeV is obtained. 
 
\section*{Acknowledgements}

We particularly wish to thank the SL Division for the efficient operation
of the LEP accelerator at all energies
 and for their continuing close cooperation with
our experimental group.  We thank our colleagues from CEA, DAPNIA/SPP,
CE-Saclay for their efforts over the years on the time-of-flight and trigger
systems which we continue to use.  In addition to the support staff at our own
institutions we are pleased to acknowledge the  \\
Department of Energy, USA, \\
National Science Foundation, USA, \\
Particle Physics and Astronomy Research Council, UK, \\
Natural Sciences and Engineering Research Council, Canada, \\
Israel Science Foundation, administered by the Israel
Academy of Science and Humanities, \\
Minerva Gesellschaft, \\
Benoziyo Center for High Energy Physics,\\
Japanese Ministry of Education, Science and Culture (the
Monbusho) and a grant under the Monbusho International
Science Research Program,\\
Japanese Society for the Promotion of Science (JSPS),\\
German Israeli Bi-national Science Foundation (GIF), \\
Bundesministerium f\"ur Bildung, Wissenschaft,
Forschung und Technologie, Germany, \\
National Research Council of Canada, \\
Research Corporation, USA,\\
Hungarian Foundation for Scientific Research, OTKA T-016660, 
T023793 and OTKA F-023259.\\


\newpage 
\begin{figure}[t]
\vspace*{-15.mm}
\begin{center}\mbox{
\epsfig{file=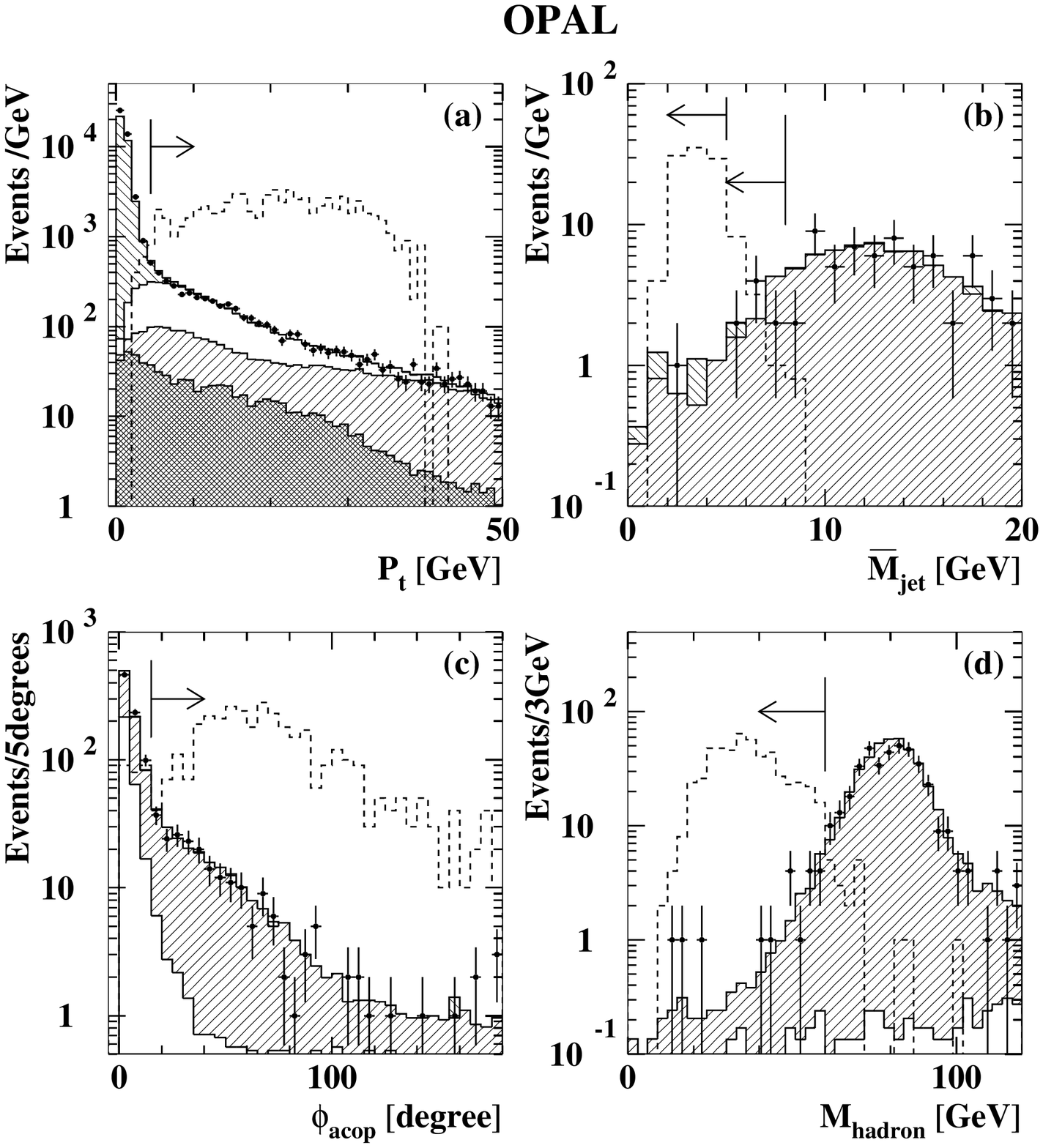,width=18.0cm}
}\end{center}
\vspace*{-3.mm}
\caption[]
{
Distributions of 
(a) ${P}_{t}$ before cut (A1), 
(b) $\mjet$ before cut (A5), 
(c) $\phiacop$ before cut (B-L3), 
(d) invariant mass excluding the most energetic lepton before cut (B-H5),
for the data, simulated background events and 
typical $\stoppair$ predictions.
In these figures, the distribution of the data is shown as points 
with error bars.
The background processes are as follows: 
dilepton events (cross-hatched area), 
two-photon processes (negative slope hatched area), 
four-fermion processes (positive slope hatched area),
and multihadronic events (open area)\@.
The arrows show the cut positions.
In (b), the left (right) arrow indicates the cut position for
$\Mvis>$65~GeV ($\Mvis<$65~GeV)\@.
The predictions for $\stoppair$ signals 
are shown by the dashed lines.
The $\stoppair$ predictions show the cases of  
($\mstop$, $m_{\neutralino}$)=(90~GeV, 70~GeV) in (a) and (b),
($\mstop$, $m_{\snu}$)=(90~GeV, 80~GeV) in (c), 
and 
($\mstop$, $m_{\snu}$)=(90~GeV, 45~GeV) in (d)\@.
The normalisations of the $\stoppair$ predictions are arbitrary.
}
\label{fig:exfig}
\end{figure}
\newpage 
\begin{figure}[htb]
\vspace*{-30.0mm}
\begin{center}\mbox{
\epsfig{file=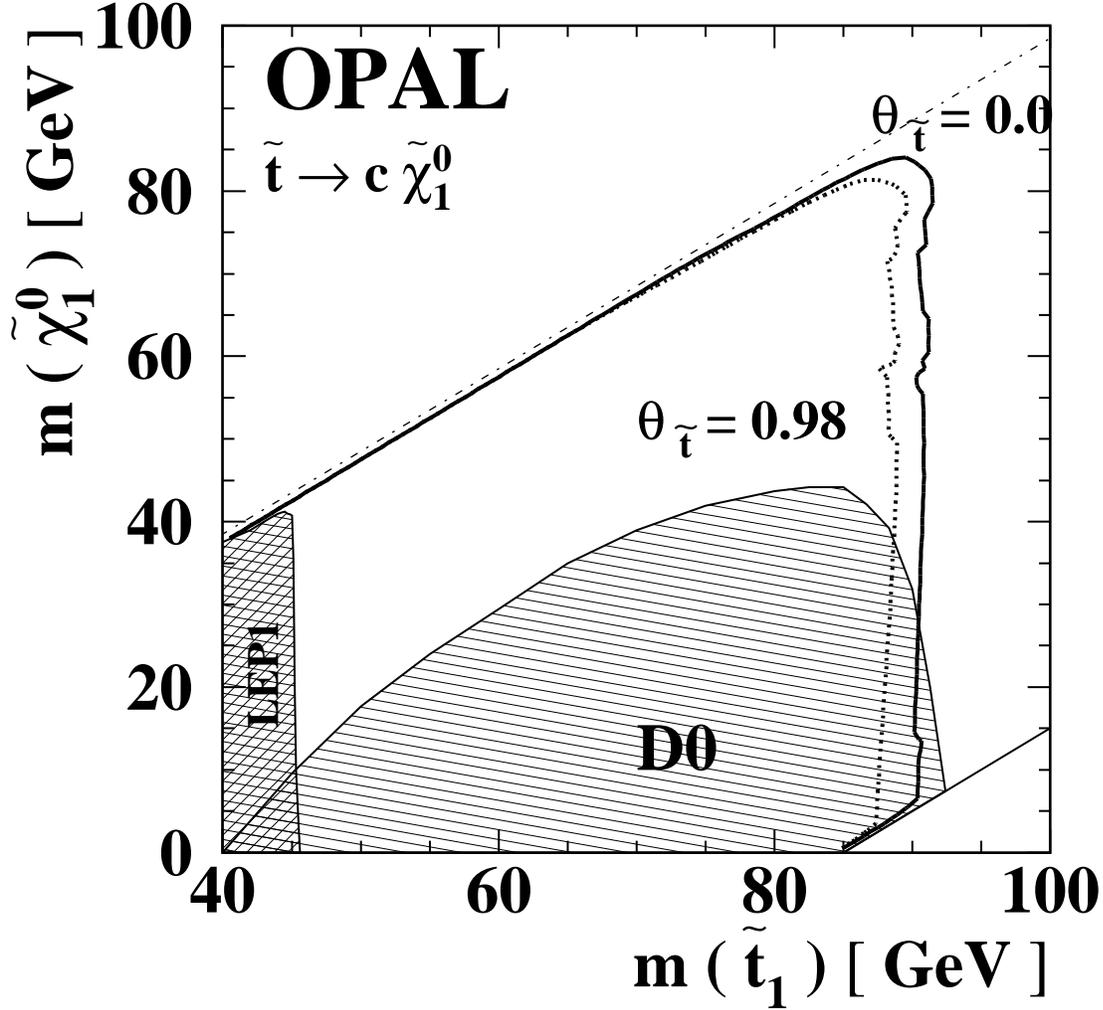,width=16.0cm}
}\end{center} 
\vspace*{-2.0mm}
\caption[]
{
The 95\% C.L. excluded regions in the $(\mstop, \, \mchi)$ plane
assuming that $\stopm$ decays into $\cq \neutralino$\@.
The solid line shows the limit for
zero mixing angle of $\stopm$, and 
the dotted line shows the limit for 
a mixing angle of 0.98~rad ($\stopm$ decouples from the $\Zboson$ boson)\@.
The cross-hatched region has already been excluded by OPAL searches 
at LEP1~\cite{opalstop}.
The singly-hatched region is excluded 
by the D0 Collaboration~\cite{d0}.
The dash-dotted straight line shows the kinematic limit 
for the $\stopm \ra \cq \neutralino$ decay.
In the triangular region of $m_{\stopm}-\mchi > m_{\Wpm} + \mb$,
the decay $\stopm \ra \bq \neutralino \Wp$(on shell)
through a virtual chargino becomes dominant.
This region is not excluded. 
}
\label{fig:result1}
\end{figure}
\newpage 
\begin{figure}[htb]
\vspace*{-15.0mm}
\begin{center}\mbox{
\epsfig{file=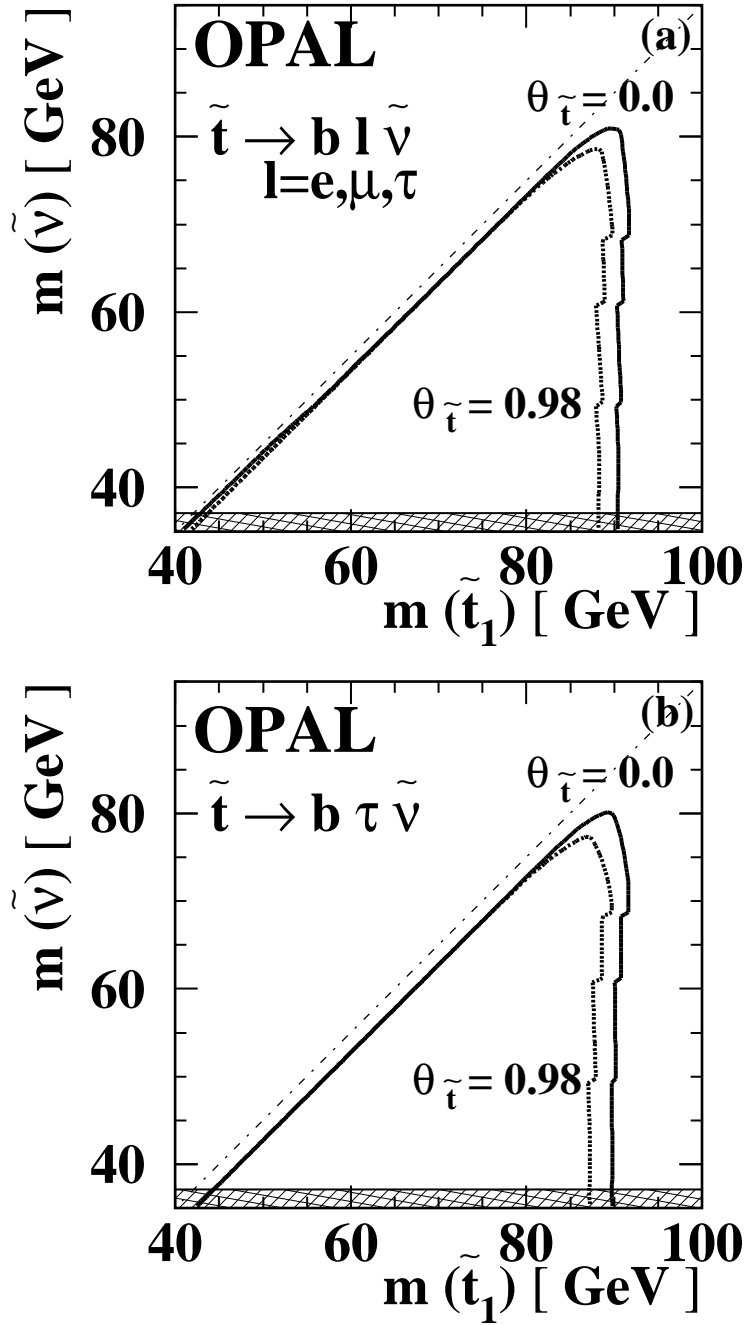,width=12.0cm}
}\end{center} 
\vspace*{-2.0mm}
\caption[]
{
The 95\% C.L. excluded regions in the $(\mstop, \, m_{\snu})$ plane
assuming that the $\stopm$ decays into $\bq \ell \snu$;
(a) the branching fraction to each lepton flavour
is the same;  
(b) $\stopm$ always decays into $\bq \tau \snu_{\tau}$\@.
The solid lines show the limits where the mixing angle of $\stopm$
is assumed to be 0.0~rad, and 
the dotted lines show the limits for a mixing angle of 0.98~rad
(decoupling case)\@.
The hatched region has been excluded at LEP1~\cite{snulimit},
and the dash-dotted diagonal line shows the kinematic limit 
for the $\stopm \ra \bq \ell \snu$ decay. 
}
\label{fig:result2}
\end{figure}
\newpage 
\begin{figure}[htb]
\begin{center}\mbox{
\epsfig{file=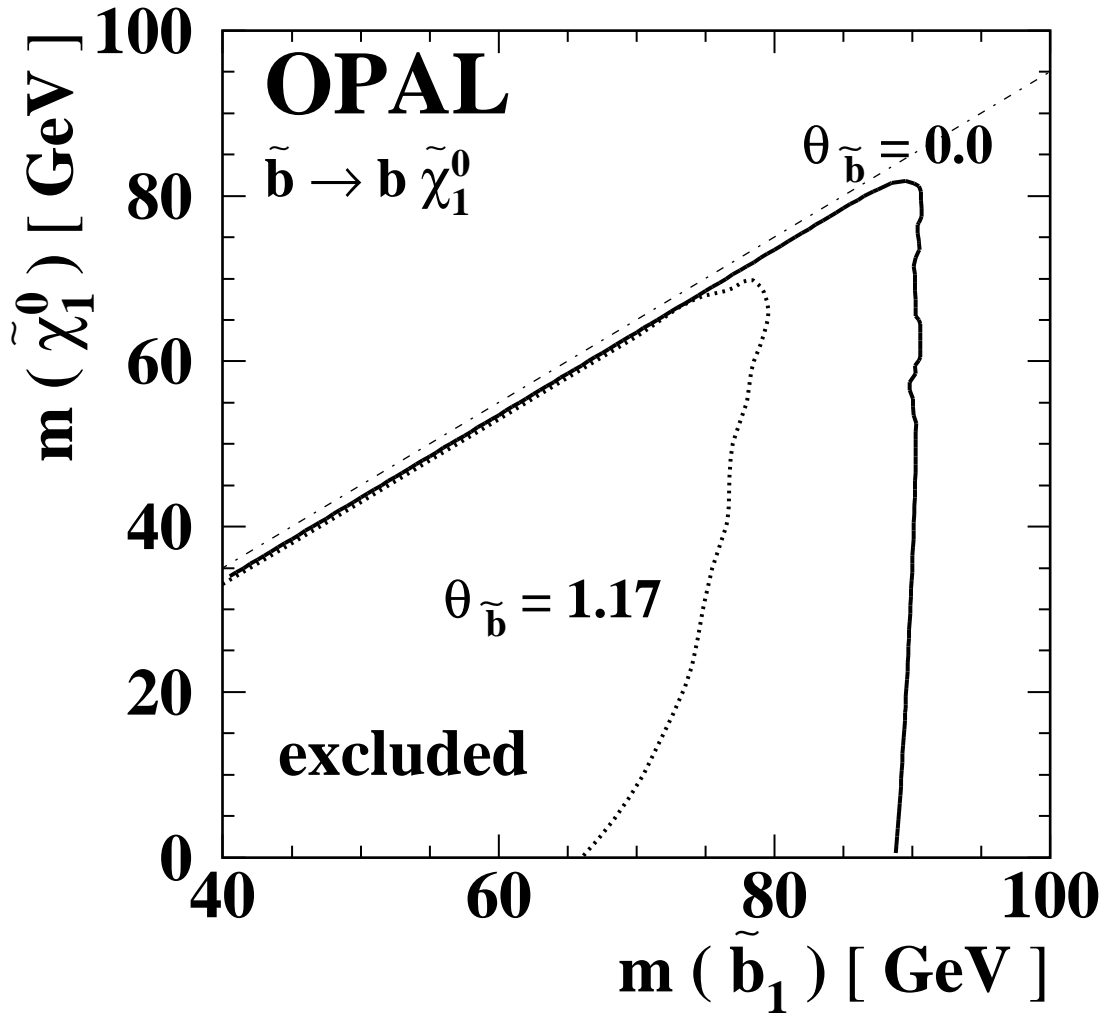,width=16.0cm}
}\end{center} 
\vspace*{-2.0mm}
\caption[]
{
The 95\% C.L. excluded regions in the $(m_{\sbotm}, \, \mchi)$ plane,
assuming 
that $\sbotm$ decays into $\bq \neutralino$\@.
The solid line shows the limit
where the mixing angle of $\sbotm$ is assumed to be 0.0~rad,
and the dotted line shows the limits for a mixing angle of 1.17~rad
($\sbotm$ decouples from the $\Zboson$ boson)\@.
}
\label{fig:result3}
\end{figure}

\end{document}